# Effect of turbine nacelle and tower on the near wake of a utility-scale wind turbine

Aliza Abraham[1,2], Teja Dasari[1,2], and Jiarong Hong[1,2]

[1]Department of Mechanical Engineering, University of Minnesota, Minneapolis, MN 55455, USA
[2]St. Anthony Falls Laboratory, University of Minnesota, Minneapolis, MN 55414, USA,

Super-large-scale particle image velocimetry (SLPIV) using natural snowfall is used to investigate the influence of nacelle and tower generated flow structures on the near-wake of a 2.5 MW wind turbine at the EOLOS field station. The analysis is based on the data collected in a field campaign on March 12$^{th}$, 2017, with a sample area of 125 m (vertical) × 70 m (streamwise) centred on the plane behind the turbine support tower. The SLPIV measurement provides the velocity field over the entire rotor span, revealing a region of accelerated flow around the hub caused by the reduction in axial induction at the blade roots. The in-plane turbulent kinetic energy field shows an increase in turbulence in the regions of large shear behind the blade tips and the hub, and a reduction in turbulence behind the tower where the large-scale turbulent structures in the boundary layer are broken up. Snow voids reveal coherent structures shed from the blades, nacelle, and tower. The hub wake meandering frequency is quantified and found to correspond to the vortex shedding frequency of an Ahmed body ($St = 0.06$). Persistent hub wake deflection is observed and shown to be connected with the turbine yaw error. In the region below the hub, strong interaction between the tower- and blade-generated structures is observed. The temporal characteristics of this interaction are quantified by the co-presence of two dominant frequencies, one corresponding to the blade vortex shedding at the blade pass frequency and the other corresponding to tower vortex shedding at $St = 0.2$. This study highlights the influence of the tower and nacelle on the behaviour of the near-wake, informing model development and elucidating the mechanisms that influence wake evolution.

## 1. Introduction

A thorough understanding of the wind turbine wake is critical for improving the efficiency of individual turbines and overall power production of a wind farm (Barthelmie *et al.* 2007; Barthelmie *et al.* 2009). In the near wake, i.e., the region within 1-4*D* downstream of a turbine, where *D* is the rotor diameter (Vermeer, Sorensen & Crespo 2003; Göçmen *et al.* 2016), coherent structures emanating from different portions of the turbine, including tip and trailing sheet vortices from the blades as well as the vortical structures from the nacelle and tower, interact strongly to affect general characteristics and stability of the wake flow and its evolution downstream (Sørensen 2011). Up to date, a large number of studies have examined in detail the behaviour of blade-generated structures, particularly the tip vortices (Widnall 1972; Okulov & Sørensen 2007; Ivanell *et al.* 2010; Sherry, Sheridan & Lo Jacono 2013; Sarmast *et al.* 2014; Nemes *et al.* 2015), but only very few studies focused on the coherent structures produced from the turbine nacelle and tower and their effect on the near-wake flows.

Nevertheless, from a series of experiments conducted by NREL on a full scale two bladed wind turbine (a diameter of 10 m, hub height of 12.2 m and a rated power of 20 kW) using the NASA Ames wind tunnel at NASA (Hand *et al.* 2001; Simms *et al.* 2001), it was revealed that the turbine tower induces significant reduction of loading (e.g. aerodynamic torque) on the blades as they pass in front of the tower. Following these experiments, several numerical studies using different simulation methods, including unsteady RANS, Reynolds averaged Navier-Stokes (Zahle, Sørensen & Johansen 2009; Wang, Zhou & Wan 2012; Li *et al.* 2012; Lynch & Smith 2014) and finite element ALE-VMS (arbitrary Lagrangian-Eulerian-variational multiscale formulation) methods (Hsu, Akkerman & Bazilevs 2014), provided further evidence of the unsteady loading effect associated with the presence of the turbine tower. In particular, using RANS on a downwind turbine, Zahle *et al.* (2009) captured the tower wake and its interaction with the rotor blades, which introduces highly transient loading on the blade. The study further noted that the axial induction caused by the rotor can considerably alter the tower vortex shedding and at times introduced a state of lock-in, a phenomenon where the natural frequency of a system is altered and synchronized with a forcing frequency, i.e. the blade passage frequency in this case. However, none of these numerical studies have systematically characterized the behaviors of these flow structures (tower and hub structures) and their effect on the near wake velocity and turbulence. In a water tunnel flow visualization experiment on a marine propeller model, Felli, Camussi & Di Felice (2011) observed that the vortical structure produced from the rotor hub, a hub vortex, meanders at a frequency equal to the rotational frequency of the rotor. Such meandering behaviour of the hub vortex was also observed from hot wire measurements on a model turbine in a wind tunnel by Iungo *et al.* (2013). However, the meandering frequency was found to be one third of the rotor frequency, consistent with the stability analysis conducted in their study. Viola *et al.* (2014) further improved on the stability analysis by incorporating effects of the Reynolds stresses through eddy-viscosity models, and provided further evidence for the meandering frequency obtained in Iungo *et al.* (2013). Applying particle image velocimetry (PIV) along a wall-parallel plane on a model wind turbine wake, Howard *et al.* (2015) also investigated the meandering motion of the hub vortex. The hub vortex signature was identified as a two-dimensional meandering line connecting the velocity minima at each streamwise location downwind of the turbine. They showed that the wavelength and amplitude of the meandering line vary with wake velocity representing a weak compressing or stretching mechanism, and are also correlated to the turbine loading. They also observed a second, lower far wake meandering frequency attributed to the rotor blockage effect. Following the same method of identifying hub vortex meandering as employed by Howard *et al.* (2015), Foti *et al.* (2016) conducted large eddy simulations (LES) on a miniature turbine and showed that the hub vortex exhibits a slow precessing motion immediately downstream of the nacelle opposite to the turbine rotation direction, resembling a spiral vortex breakdown. The helical meandering motion of the hub vortex further downstream was found to arise from this precessing motion. Using the actuator disk, actuator line and turbine resolving LES techniques on a hydrokinetic turbine, Kang, Yang & Sotiropoulos (2014) conducted a systematic investigation of the wake meandering and pointed out that the interaction between turbine hub and blade-tip vortices in the near wake triggers wake meandering.

Only recently, a number of numerical studies have started looking into the effect of nacelle and tower on wind turbine wake flows. Specifically, using novel actuator surface models for the turbine blades and nacelle, Yang & Sotiropoulos (2018) showed that the turbine nacelle increases turbulent kinetic energy (TKE) significantly in the near wake. Wang *et al.* (2017) conducted LES on model-scale turbines with fully-resolved turbine geometry using an immersed boundary method, and demonstrated the pronounced effect of both nacelle and tower on the near wake characteristics. Specifically, the nacelle and tower were found to generate substantial amounts of turbulence, which in turn enhanced downstream flow mixing. However, the authors suggested that the pronounced effects of nacelle and tower are caused by the relatively larger nacelle and tower sizes of the model turbines employed in the study in comparison to those in utility-scale wind turbines. Similarly, Santoni *et al.* (2017) applied LES to a scaled-down turbine with well-resolved tower and nacelle geometries, and showed that the tower and nacelle cause a substantial velocity deficit and affect TKE and associated fluxes, especially in the portion of the wake with tower influence. The tower wake was also found to interfere with the turbine blade wake, promoting tip vortex breakdown.

Despite these recent studies mentioned above, no field study has ever been performed on utility-scale wind turbines to investigate the behavior of the flow structures generated from turbine hub and nacelle and determine their effect on near-wake characteristics. Considering the significant discrepancy between field and laboratory conditions, including Reynolds number, atmospheric condition, turbine characteristics, etc., such a study is necessary for validating the observations derived from laboratory and numerical work, and can potentially reveal new and important physics occurring at utility-scale settings. Needless to say, the measurements required for such study are very difficult to obtain with conventional field measurement tools (e.g., lidar, sodar and radar, etc.), which do not yield sufficient spatial and temporal resolution to effectively capture dynamics of near-wake coherent structures and their interactions.

Nonetheless, by taking advantage of natural snowflakes as flow tracers, a recent study by Toloui *et al.* (2014) first introduced super-large-scale particle image velocimetry (SLPIV) for flow measurements in the atmospheric boundary layer (ABL) over large fields of view. Specifically, they implemented SLPIV to characterize the ABL through a pilot experiment in an area of ~ 22 m × 52 m with a spatial resolution of 0.34 m and a temporal resolution of 15 Hz, which was validated by comparing with sonic anemometer measurements from a meteorological tower nearby. Successively, Hong *et al.* (2014) applied the same technique along with flow visualization to study the complex flow field and coherent structures in the near wake of a 2.5 MW wind turbine. This work quantified a flow field of ~ 36 m × 36 m at ~ 0.3$D$ downstream of the turbine, and further demonstrated correlations between the turbine operation and the coherent structures in the near wake. Moreover, Dasari *et al.* (2019) further implemented SLPIV to quantify a larger flow field 115 m × 66 m at ~ 0.4$D$ downstream of the turbine and 0.19$D$ offset from the central tower plane and compared the wake velocity deficit with a number of existing wake models. Owing to high spatiotemporal resolution of SLPIV, the study revealed an intermittent wake contraction behaviour (about 25% of time) which was found to be correlated with the rate of change of blade pitch. In addition, the study showed correlations between the tip vortex behavioural patterns and turbine operation & response characteristics like power, tower strain, blade pitch, angle of attack and their fluctuations which opens up avenues to predict the wake behaviour based on readily available turbine operational parameters. However, as the study was conducted on a plane offset from the tower plane, the coherent structures emanating from the nacelle and tower could not be visualized and their potential implications on the near wake could not be studied.

Following Dasari *et al.* (2019) and using similar flow visualization and SLPIV techniques with natural snowflakes, the current paper focuses on investigating the flow structures produced by the tower and nacelle of a utility-scale turbine and their impact on near-wake mean flow and turbulence characteristics. The study is based on a field campaign conducted in March 2017 at the symmetry plane behind the tower in the near wake of the 2.5 MW turbine used in Hong *et al.* (2014) and Dasari *et al.* (2019). The paper is structured as follows: § 2 provides a brief description of the experimental methods, field campaign and data processing procedures. § 3.1 reports the results of the near-wake velocity field captured at the turbine symmetry plane and spanning from the bottom-blade tip to the top tip. § 3.2 focuses on the dynamic behaviours of the coherent structures shed from the nacelle and tower. § 4 provides a summary and discussions of the results.

## 2. Experimental Method

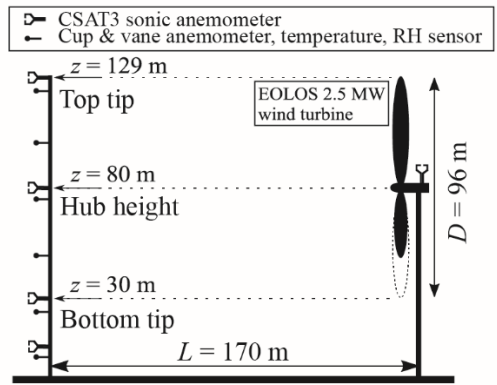

FIGURE 1. Schematic of the 2.5 MW turbine and met tower at the EOLOS station.

The field experiments are conducted at the EOLOS Wind Energy Research Field Station (referred to as the EOLOS station hereafter) in Rosemount, MN. As shown in figure 1, the EOLOS station consists of a 2.5 MW Clipper Liberty C96 wind turbine (referred to as the EOLOS turbine hereafter) and a 130 m meteorological tower (referred to as the met tower hereafter). The EOLOS turbine is a 3-bladed, horizontal-axis, pitch-regulated machine with a rotor diameter ($D$) of 96 m and a supporting tower 80 m in height, capable of operating at variable wind speed. The turbine tower has a cylindrical cross section with a constant diameter of 4.1 m up to a height of 26 m and a conical cross section above, tapering gradually to a diameter of 3 m at 78 m elevation. The nacelle of the turbine is near cuboidal in shape with dimensions of 5.3 m × 4.7 m × 5.5 m. The experiments employ the SLPIV and flow visualization technique from Hong *et al.* (2014) & Dasari *et al.* (2019) to investigate the flow field and the coherent structures in the near wake. Briefly, the experimental setup is composed of an optical assembly for illumination, a camera and the corresponding data acquisition system.

The optical assembly includes a 5 kW highly collimated search light and a curved reflecting mirror for projecting a horizontal cylindrical beam into a vertical light sheet. The camera, Sony-A7RII mounted with a 50 mm $f$/1.2 lens, is used to provide 4K-resolution, 30 Hz videos for video recording. Please refer to Dasari *et al.* (2019) for further information on the turbine, associated instrumentation and experimentation technique.

The present paper studies 1 hour of data obtained from a deployment conducted on March 12$^{th}$, 2017 in the wake of the EOLOS turbine. The details regarding the dataset are summarized in tables 1 and 2 with a schematic presented in figure 2 illustrating the key parameters for the measurement setup. Specifically, as shown in table 1 and figure 2, the location of the field of view (FOV) for each dataset is characterized by its downstream distance from the tower ($x_{FOV}$), its offset from the central tower plane ($y_{FOV}$) and the elevation of the FOV center above the ground level ($z_{FOV}$). The dimension of the FOV is characterized by its height ($H$) and width ($W$). The distance between the camera and the light sheet is represented by the $L_{CL}$ with $\theta$ indicating the tilt angle of the camera. The FOV is located very close to the central tower plane (0.06$D$ offset from the tower plane). This dataset is exclusively used for quantitative examination of the near-wake turbulent flow field. Table 2 summarizes the detailed meteorological information as well as the turbine operational conditions. The wind speed and direction are measured by the sonic anemometer at the nacelle of the EOLOS turbine. Temperature and humidity are measured at the met tower. The turbulence intensity and Obukhov length are calculated from the acquired data accordingly.

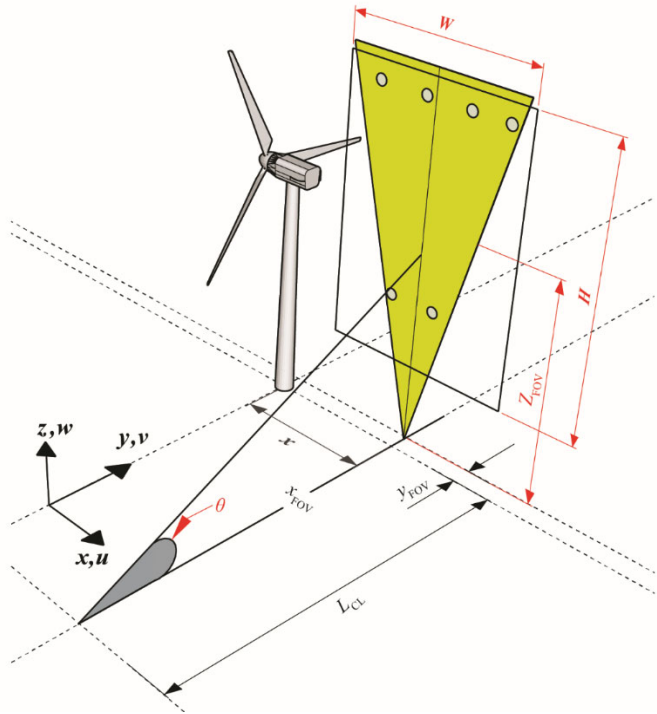

FIGURE 2. Schematic of the measurement setup used in the deployment.

| Deployment Date | Dataset Duration | FOV Location ($x_{FoV}$, $y_{FoV}$) | FOV Elevation ($z_{FoV}$) | FOV Size ($H \times W$) | Camera-to-light Distance ($L_{CL}$) | Tilt angle ($\Theta$) |
|---|---|---|---|---|---|---|
| March 12$^{th}$, 2017 | 62 min | 0.35$D$, 0.06$D$ | 80 m | 125 m × 70 m | 171 m | 24.5° |

TABLE 1. A summary of the key parameters of the measurement setup for the deployment dataset used in the present paper.

| Deployment dataset | Mean wind speed at hub height | Mean wind direction (from North) | Turbulence intensity | Mean temperature | Relative humidity | Obukhov Length | Turbine operational region | Tip-speed-ratio |
|---|---|---|---|---|---|---|---|---|
| March 12th, 2017 | 5.9 m/s | 58° | 0.18 | - 8.1 °C | 93 | 131 | 1.5 – 2 | 8 – 11.5 |

TABLE 2. A summary of the key parameters of the atmospheric and turbine operational conditions.

It is to be noted that conditions related to the experimental setup orientation (viz. wind direction, turbine nacelle direction) change continuously during the deployment. Figure 3 provides a time series of these parameters during the night of March 12th, 2017. The figure also showcases the 3 data collection periods, which together constitute 62 minutes of visualization data. From this plot it is clear that there is considerable misalignment up to 25° between the light sheet direction (dashed line in the figure) and the wind direction (solid line) during this period. Such misalignment results in up to 9% reduction in the mean velocity magnitude measured at the tower plane, and more importantly, a disappearance of hub and tower flow structures in the sample plane at high degrees of misalignment. To account for this, conditional sampling is applied based on the misalignment angle, $Y_{LW}$, which is the magnitude of the difference between the wind direction and the light sheet direction. When $Y_{LW} \leq 10°$, the data is considered aligned, and when $Y_{LW} \geq 20°$, the data is considered misaligned. These cut-off values were selected to ensure sufficient data in each category (600 seconds and 16% of the data is included in each category).

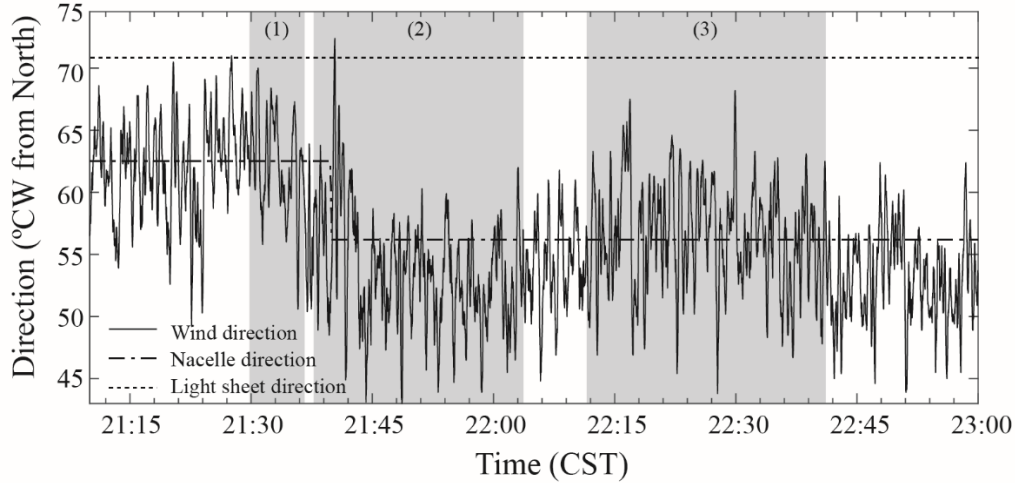

FIGURE 3 Time series of wind, nacelle, and light sheet direction on March 12th, 2017 marked with video data collection periods.

Figure 4 shows the series of image pre-processing steps undertaken to enhance the images before applying PIV analysis. The images in this dataset were procured with guy wires supporting the met-tower in the FOV as shown in figure 4(*a*). These support cables are removed through a spatial filtering process in the Fourier domain, rendering the image shown in figure 4(*b*) which is then de-warped (figure 4*c*), and further enhanced through sliding background subtraction to obtain final image shown in figure 4(*d*). The de-warping process is similar to the one employed in Dasari *et al.* (2019).

The velocity vectors are calculated using the adaptive multi-pass cross correlation algorithm from *LaVision Davis 8*. Due to the insufficient resolution of the images owing to large viewing distances, the large scale patterns formed by the snowflakes and coherent structures in the near wake are tracked to quantify the flow field (Dasari *et al.* 2019). The cross correlation was first conducted using an initial interrogation window of 128 × 128 pixels which was then reduced to 32 × 32 pixels with 50% overlap, providing a spatial resolution of 4.2 m/vector. The cross correlation is also applied to image pairs with 5 frame skip in a time sequence of images to ensure sufficient displacement of snow patterns between the two images within a pair, resulting in temporal resolution of 6 Hz.

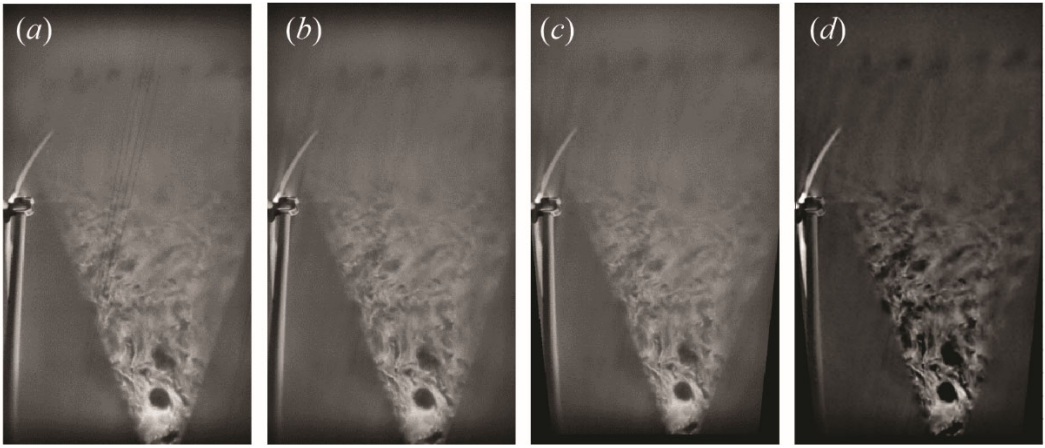

FIGURE 4. A sample of (*a*) raw image, (*b*) Fourier filtered image (*c*) de-warped image and (*d*) enhanced image after background subtraction showing the steps of image processing before the velocity vector calculation.

## 3. Results

### 3.1. *Mean Wake Statistics*

The near-wake flow in the tower plane is first examined using the mean velocity field. Figure 5 presents the time average of the velocity field for the aligned (figure 5*a*) and misaligned data (figure 5*b*). Figure 5(*c*) compares the velocity deficit profile to the velocity deficit profile at a different spanwise location but similar streamwise location from Dasari *et al.* (2019). The horizontal dashed lines indicate the location of the top blade tip, the hub, and the bottom blade tip. Remarkably, the aligned velocity field highlights a prominent high-speed flow region behind the rotor hub in comparison to the flow with expected velocity deficit behind the blades. According to Magnusson (1999), this near-wake flow pattern is associated with a reduction in lift, and correspondingly axial induction, at the blade roots. This accelerated flow region has been observed in wind tunnel studies where the model blade geometries cause a similar reduction in lift at the roots (Hancock & Pascheke 2014, Foti *et al.* 2018), and in LES studies where the turbine geometry accurately represents a field-scale turbine (Schulz *et al.* 2017). Gallacher & More (2014) observed this phenomenon in the average sense using nacelle-mounted lidar on an offshore wind turbine. The mean flow at the tower plane also shows an increased velocity deficit behind the turbine tower caused by additional tower-induced blockage, as previously observed in a recent numerical simulation studying the effect of the tower on the wake flow (Santoni *et al.* 2017). Moreover, a flow acceleration region is observed below the bottom tip (evidenced by the higher magnitude velocity in this region compared to the region above the top tip) due to rotor blockage, confirming the observation from the field studies of Iungo, Wu & Porté-Agel (2012) using lidar and Hong *et al.* (2014) based on snow PIV. By comparing the aligned velocity field at $x = 10\ m$ to that at $x = 60\ m$, particularly in the region of $z = 110\ m$, a slight increase in velocity deficit in the downstream direction is observed. This phenomenon has been previously observed in field scale studies using lidar (Käsler *et al.* 2010, Gallacher & More 2014).

In the misaligned velocity field shown in figure 5(*b*), the high-speed region behind the hub only exists directly behind the turbine and reduces quickly in the spanwise direction, exiting the imaging plane 30 m downstream of the turbine. With a known average misalignment angle of $Y_{LW} = 22°$, the spanwise extent of the accelerated flow region can be estimated as 11 m on either side of the hub, approximately twice the width of the nacelle and $0.1D$. Comparing the velocity deficit profile with an off-symmetry plane at a similar downstream location (figure 5*c*) further reinforces the limited extent of the hub wake. The tower plane velocity deficit profile is taken at $x/D = 0.41$ and calculated as $\Delta U(z)/U_\infty = 1 - U(z)/U_\infty(z)$, where $U_\infty(z)$ is the incoming boundary layer profile measured by the sonic anemometers on the met tower. The off-tower plane profile is measured at $y/D = 0.19$, only 12 m away from the tower plane, and the acceleration around the hub height is significantly reduced. In the wake symmetry plane the velocity deficit profile demonstrates a strong double-Gaussian shape, providing evidence for the analytical model for the near-wake proposed by Keane *et al.*

(2016). The tower plane profile also exhibits larger velocity deficits than the off-tower plane above and below the hub, with local velocity minima at the points of maximum blade thrust, and increased vertical asymmetry due to the increase in velocity deficit caused by the tower.

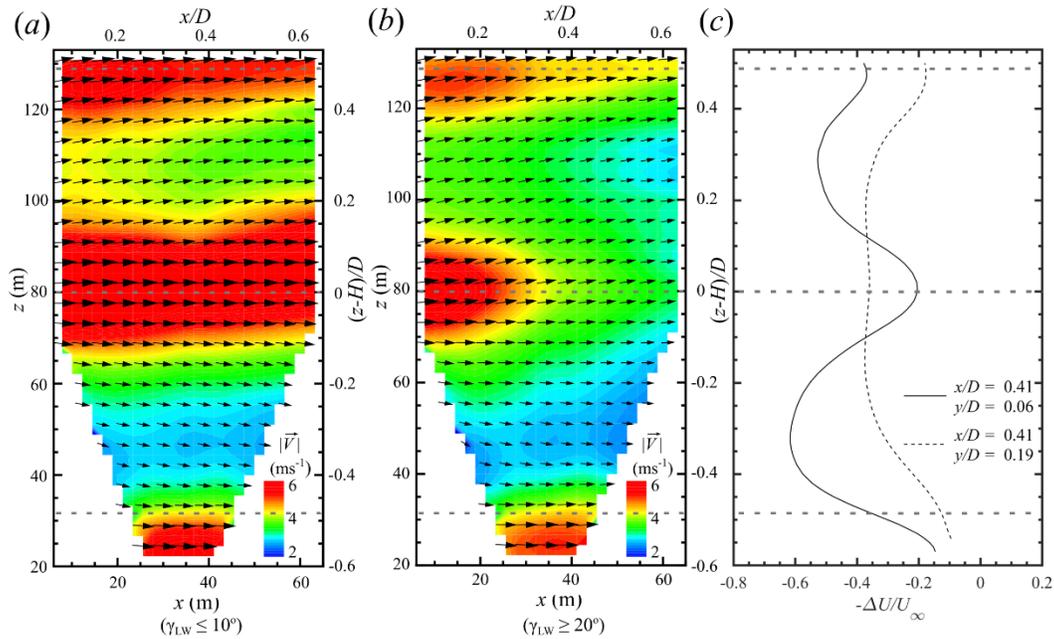

FIGURE 5 (*a*) Time-averaged velocity vector field (1:2 skip applied in horizontal and vertical directions for clarity) superimposed with the velocity magnitude contours at the tower plane under aligned ($\gamma_{LW} \leq 10°$), and (*b*) misaligned conditions ($\gamma_{LW} \geq 20°$). (*c*) Comparison of the tower plane velocity deficit profile under the aligned condition with the off-tower plane ($x/D = 0.41$) measurement presented in Dasari *et al.* (2019).

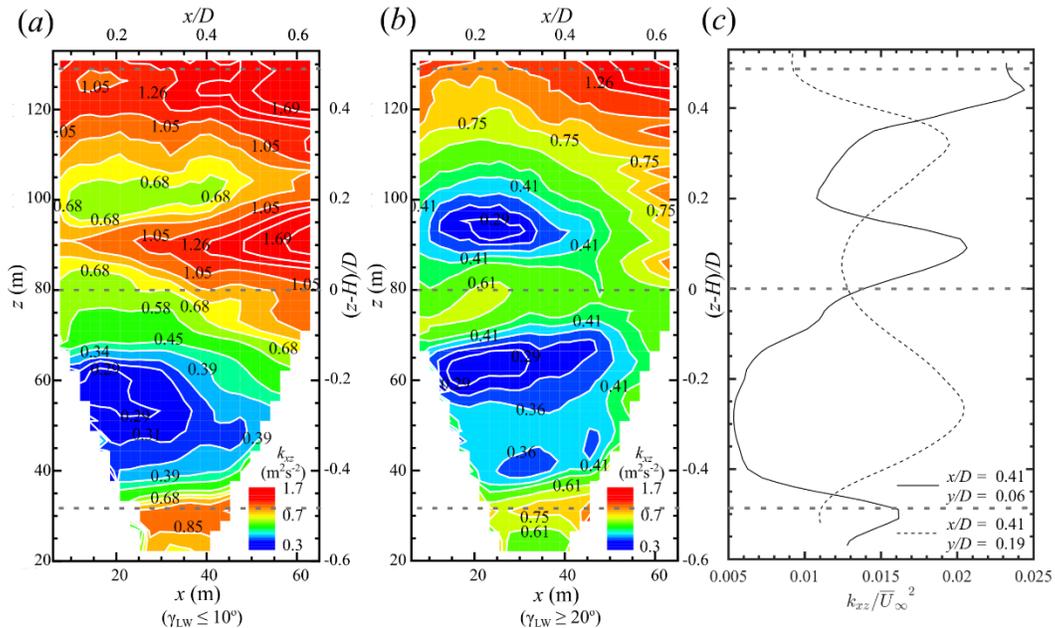

FIGURE 6 (*a*) In-plane TKE contours at the tower plane under aligned ($\gamma_{LW} \leq 10°$), and (*b*) misaligned conditions ($\gamma_{LW} \geq 20°$). (*c*) Comparison of the tower plane TKE profile under the aligned condition with the off-tower plane ($x/D = 0.41$) measurement presented in Dasari *et al.* (2019).

Figure 6 shows the in-plane turbulent kinetic energy (TKE) field, i.e. $k_{xz} = \frac{1}{2}(\langle u'^2 \rangle + \langle w'^2 \rangle)$ for the (*a*) aligned and (*b*) misaligned data, while (*c*) shows the in-plane TKE profile at the symmetry plane compared with the spanwise location from Dasari *et al*. (2019). The aligned in-plane TKE demonstrates high values in regions of high shear, i.e., behind the blade tips and the nacelle. These TKE peaks have also been observed in LES studies (Wu & Porté-Agel 2012, Yang *et al*. 2015). The TKE near the top blade tip is higher than that near the bottom tip because there is higher shear between the rotor wake and the free stream flow at the top of the rotor (Chamorro & Porté-Agel 2009, Wu & Porté-Agel 2012). Interestingly, the peak in TKE in the region of the nacelle occurs slightly above the hub height. The offset of this peak is likely due to slight light sheet misalignment with the center of the wake, even for the conditionally sampled aligned case. The average misalignment angle of the aligned data is $Y_{LW} = 7°$, indicating the light sheet is intersecting with the wake at a slightly off-center location. Using wind tunnel experiments on laboratory-scale models, Chamorro & Porté-Agel (2010) and Zhang, Markfort & Porté-Agel (2012) have shown that wake rotation causes asymmetry in the turbulence distribution of the turbine near wake. Because of the slight offset, the light sheet intersects the wake at a location where the flow containing increased TKE has been deflected upwards by the wake rotation. In the misaligned case (figure 6*b*), the hub TKE peak is significantly reduced, reinforcing the limited spanwise range of the hub wake influence. From the comparison between the in-plane TKE profile from the tower plane taken at $x = 0.41D$ and the profile from the off-tower plane at the same streamwise location (figure 6*c*), it is clear that the hub TKE peak completely disappears in the off-tower plane. This is consistent with the diminishing hub wake signature in the off-tower plane shown in the velocity deficit profiles in figure 4(*c*).

A strong reduction in TKE behind the tower is also apparent in the symmetry plane. Similar observations were made in laboratory-scale (Chamorro & Porté-Agel 2009) and LES studies (Wu & Porté-Agel 2012), but the phenomenon has not been explained in detail. Here we would like to attribute this reduction in TKE to the effect of the tower breaking up large-scale streamwise turbulent structures in the atmospheric boundary layer, interrupting the turbulence cycle. This explanation is consistent with the literature on the effect of a wall perturbation on a turbulent boundary layer (Hamilton, Kim & Waleffe 1995; Pearson, Elavarasan & Antonia 1998; Jacobi & McKeon 2011; Ryan, Ortiz-Dueñas & Longmire 2011; Pathikonda 2013; Pathikonda & Christensen 2015). Because the turbulence cycle is responsible for most of the TKE in a turbulent boundary layer, the interruption of this cycle causes a reduction in TKE (Jiménez & Pinelli 1999). In addition, it is observed that, though figure 6 shows a reduction of in-plane TKE behind the midspan of the tower, an increase is apparent near the bottom and top of the tower. The increase at the top of the tower (above 60 m) occurs because the large-scale near-wall energy is deflected upwards above the tower in response to the lower pressure above the cylinder in the boundary layer, causing flow reattachment and a corresponding peak in turbulence at the free end. The increase in TKE at the base of the tower near the wall (below 40 m) is caused by the base vortex formed behind the bottom of the tower by the interaction between the downwash flow from the free end of the cylinder and the upwash flow from the wall. This explanation is consistent with the literature on large aspect ratio cylinders immersed in boundary layers (Castro & Robins 1977; Park & Lee 2002; Sumner, Heseltine & Dansereau 2004; Stoesser *et al*. 2010; Krajnović 2011; Jacobi & McKeon 2011).

### 3.2. *Dynamic Behaviour of Coherent Structures*

Coherent structures in the near wake are visualized as regions of low snowflake concentration, or voids. These voids are the result of particle inertia in regions of high vorticity in the turbulent flow, as described in detail by Hong *et al*. (2014) and Dasari *et al*. (2019). Previous studies have observed blade tip vortices and trailing sheet vortices, but the alignment of the field of view in the current study with the tower plane enables the visualization of the vortices shed from the blade roots, nacelle, and tower. A sample video frame exhibiting all of these structures is shown in figure 7. The most apparent feature of this image is the contrast between the bottom and top halves of the wake. The top half (above the hub) shows regular tip, root, and trailing sheet vortices shed periodically from the turbine blades. In the bottom half, however, the structures are much more chaotic as a result of the interaction between vortices shed from the blades and those shed from the tower. In the region behind the tower, tip and sheet vortices are still visible, but they are heavily distorted. The hub vortex, distinct from the blade root vortices, is visible in the region directly behind the hub.

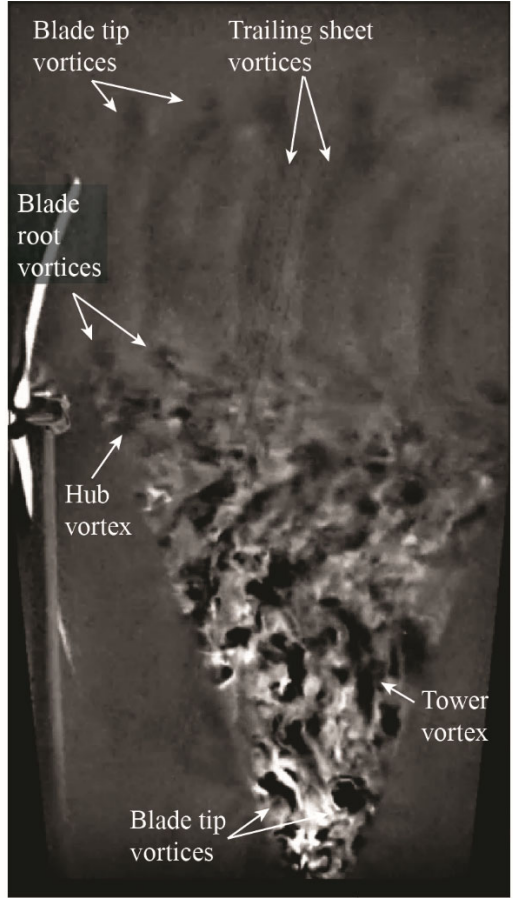

FIGURE 7 Coherent structures in the near wake of the turbine, visualized as regions of snowflake voids.

The dynamic characteristics of the hub wake are first investigated using the method introduced in Howard *et al.* (2015) and Foti *et al.* (2016) for quantifying the wavelength of hub wake meandering. The steps of this process are shown in figure 8. A region of the velocity profile around the hub between $-0.2 < (z - H)/D < 0.2$ at a downstream location of $x/D = 0.41$, the same downstream location as the profiles in figures 5 and 6, is selected (figure 8*a*). The elevation corresponding to the center of the hub wake is identified for each time step as the local velocity maximum within this region, defined as $z_{\text{HW}}$ (figure 8*b*), and Taylor's hypothesis using the mean convection velocity in the nacelle region of the wake is applied to convert the time sequence to a spatial meandering signature. A sample sequence of the hub wake center position is shown as the light gray line in figure 8(*c*). Note that Howard *et al.* (2015) and Foti *et al.* (2016) defined the hub wake center as the velocity minimum because they did not observe a region of flow acceleration around the hub. This difference may be due to the fact that both studies were conducted on laboratory scale model turbines where the blade profile and the nacelle and rotor diameter ratio did not match the utility-scale turbine. The hub wake position sequence is then low-pass filtered with a cutoff frequency equal to the average hub rotation frequency to focus the analysis on larger-scale behaviors. The filtered signal is the black line in figure 8(*c*). The wavelength is calculated as the distance between adjacent peaks and adjacent troughs, as shown in figure 8(*c*). Wavelengths from the entire dataset are compiled into a probability distribution function (PDF) and fit with a normal distribution using the method of maximum likelihood estimation, providing a mean wavelength of $\lambda/D = 0.55$ and a standard deviation of 0.19 (figure 8*d*). This mean value is consistent with that observed by Foti *et al.* (2016) in the extreme near wake of a model turbine ($x < D$). Calculating a Strouhal number for hub wake meandering based on the rotor diameter leads to $St_\text{D} = \frac{fD}{U_{hub}} = \frac{U_c D}{\lambda U_{hub}} = 1.7$, a value significantly higher than that observed by previous laboratory-scale studies: $St_\text{D} = 0.7$ (Howard *et al.* 2015, Foti *et al.* 2016). This discrepancy could be related to the reduced velocity deficit around the hub, not observed in the aforementioned studies, which would affect the ratio $\frac{U_c}{U_{hub}}$. However, calculating the Strouhal number based on the nacelle dimension gives $St_n = 0.06$, a value consistent with Howard *et al.* (2015) who reported $St_n = 0.06$, and close to the range of $St_n = 0.03 - 0.05$, calculated from the frequency, velocity, and geometry provided by Iungo *et al.* (2013) for a different laboratory-scale model turbine. This value of $St_n$ is also consistent with studies investigating vortex shedding from an Ahmed body, a roughly cuboidal bluff body used to model automobile aerodynamics (Duell & George 1999, Krajnović & Davidson 2003). These studies have shown that vortices shed from the upper and lower surfaces of the body interact to form a ring vortex which moves back and forth, generating a low frequency pumping motion that sheds vortices from the back of the separation bubble. The Strouhal number of this pumping is in the range of 0.059 to 0.069, matching the Strouhal number observed in the current study. This suggests the hub wake behavior is influenced by both the rotor dynamics and bluff body shedding from the nacelle. The rotor generates the region of accelerated flow around the hub, and the nacelle sheds coherent structures at the frequency governed by the interaction between vortices generated by the shear on its surfaces. Both of these effects are difficult to model on the laboratory scale due to discrepancies in blade shape and relative nacelle size compared to utility-scale turbines.

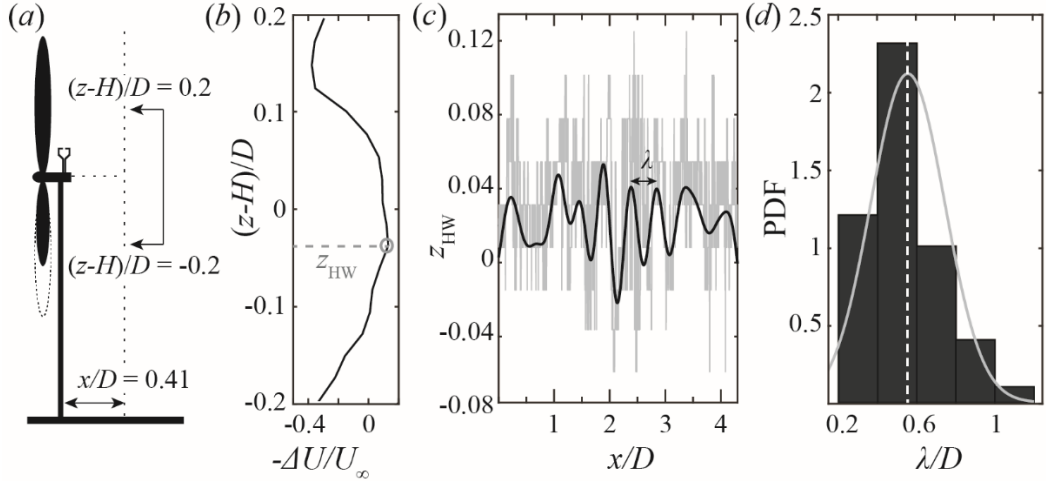

FIGURE 8 (*a*) Schematic illustrating the region of the velocity profile selected for the investigation of hub wake meandering. (*b*) Sample velocity profile section with the location of the hub wake center, $z_{HW}$, defined by the maximum velocity within this region, indicated by a gray circle and a horizontal dashed line. (*c*) A sample unfiltered (light gray) and filtered (black) sequence of hub wake position as a function of downstream distance as calculated using Taylor's hypothesis. The distance between adjacent peaks and troughs defines the local wavelength, $\lambda$. (*d*) The PDF of the hub wake meandering wavelength over the entire dataset, fitted with a normal distribution (light gray line). The mean value is $\lambda/D = 0.55$, indicated by a vertical dashed line, and the standard deviation is 0.19.

In addition, as shown in figure 9, the entire hub wake has been observed to shift upward and downward for persistent periods of time during our measurement. This phenomenon, referred to as wake deflection, is characterized using the time variation of $z_{HW}$, filtered using a 30-second moving average (the filtered signal is defined as $\tilde{z}_{HW}$). The filter is applied to remove the meandering effect described above and to reflect the persistence of the deflection throughout the streamwise span of the FOV. Figure 10 shows the relationship between hub wake deflection and turbine yaw error, defined as the difference between incoming wind direction and nacelle direction. The hub wake deflection is positively correlated with yaw error, with a correlation value of $R = 0.3$ (figure 10*a*). The correlation is weakened by constantly changing conditions in the field, i.e., incoming wind speed, misalignment angle, incoming turbulence, etc. However, when conditions are stable, the correlation is clearly visible in a sample time series (figure 10*b*). In the three-minute sequence in figure 10(*b*), the solid line represents the smoothed vertical hub wake position, defined by the location of the velocity maximum as in figure 8(*b*). When this is greater than 0, the hub wake is above the nacelle, and when this is less than 0, the hub wake is below the nacelle. The dashed line in figure 10(*b*) represents the yaw error. The relationship between yaw error and vertical hub wake position can again be explained by the nacelle's geometric resemblance to an Ahmed body. Wind tunnel studies of Ahmed bodies have shown that, under yawed conditions, the pressure distribution across the body causes the vortices shed behind the body to be deflected upwards on the leeward side and downwards on the windward side (Gohlke et al. 2007, Keogh et al. 2016). The light sheet in the current study is on the leeward side when yaw error is positive, causing an upward wake deflection, and on the windward side when yaw error is negative, causing a downward deflection.

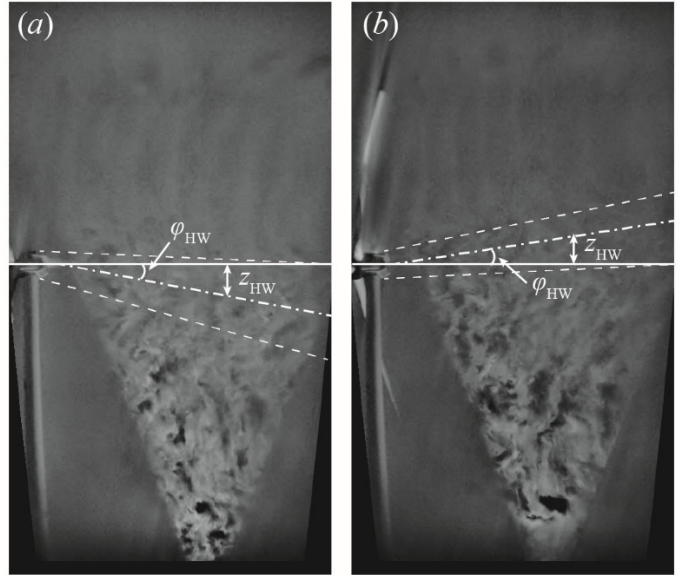

FIGURE 9 Sample snapshots of the snow void patterns in the turbine near wake showing the (a) downward and (b) upward deflection of the hub wake with respect to the elevation of the turbine nacelle. The approximate centerline (dash-dotted line) and the boundary (dashed lines) of the hub wake are highlighted in the figure, as well as the hub wake deflection angle ($\varphi_{HW}$) and the hub wake height at $x = 0.41D$ ($z_{HW}$).

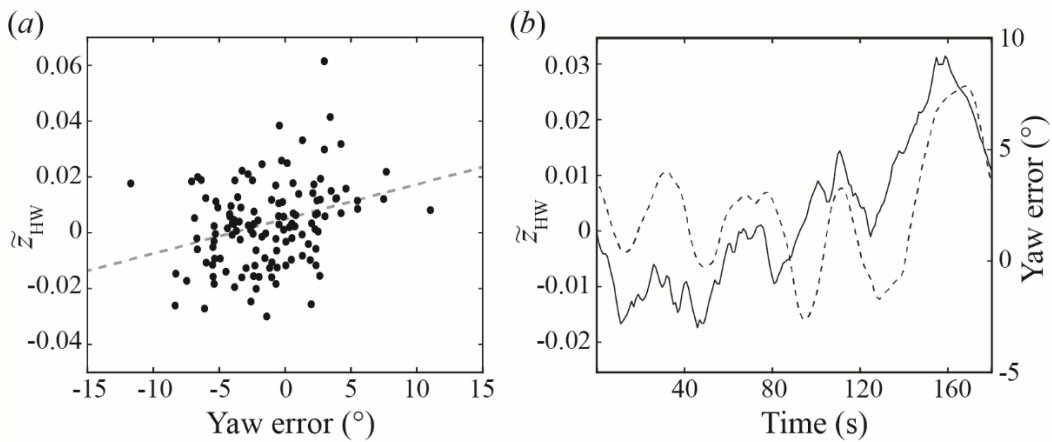

FIGURE 10 (a) Scatter plot showing the relationship between yaw error and vertical hub wake position. Each data point represents 30 seconds of data. The dashed line shows the linear regression of the data. (b) A sample three-minute time sequence showing the correspondence between hub wake position (solid line) and yaw error (dashed line).

The characteristics of coherent structures shed from the tower are investigated using spectral analysis of the streamwise velocity vectors. Premultiplied frequency spectra are calculated at three different locations in the wake, indicated by I, II, and III in figure 11(a). Location I is below the top blade tip and above the hub, II is below the hub and above the bottom tip, and III is below the bottom tip. The vectors used to calculate the spectra are all taken at $x/D = 0.41$, as in figures 5, 6, and 8. The spectra are calculated over 30-second windows to account for changing parameters such as wind speed, wind direction, hub speed, etc. Figure 11(b) shows sample premultiplied spectra for each location, with a gray circle and dashed vertical line marking the location of the most prominent peak. The peaks of the spectra for all the time windows in the dataset are combined into PDFs, shown in figure 12 for each location. Each PDF is fit with a bimodal normal distribution to clarify the trends. These distributions reveal peaks within two frequency ranges, indicated by dark and light gray bands in the figure. At the location above the hub (I), the distribution contains a single peak within the blade pass frequency band during the time period of our measurements, indicated by the light gray band and within the range of 0.52-0.74 Hz. This frequency band corresponds to a rotor speed of 10.5-14.8 RPM and tip speed ratio of 7.5-12.1. The prevalence of this frequency is caused by the structures shed from the blades in the upper half of the wake. Below the hub but above the bottom tip (II), a second frequency emerges corresponding to tower vortex shedding at a $St = 0.2$, matching the Strouhal number of the vortex shedding behind a cylinder at high Reynolds numbers (e.g. Shih *et al.* 1993). The peak of this frequency mode is in the range of 0.22-0.41 Hz, calculated based on a cylinder diameter of 4.1 m and incoming wind speeds within the range of 4.6-8.4 m/s. The signature of the blade pass frequency is still clear at this elevation, but it is significantly weakened by interactions between blade structures and tower structures. Below the bottom tip (III), the signature of the tower vortex frequency is dominant. The blade pass frequency is still visible because of blade tip vortices that are convected below the bottom tip during periods of strong wake expansion. These PDFs indicate that the presence of the tower interferes with the structures shed from the blades, reducing the occurrence of structures at the blade pass frequency. This finding provides evidence for the observation made in the simulation described by Santoni *et al.* (2017) that the tower wake interacts with the blade-shed structures, causing tip vortex breakdown. Their study emphasized the importance of including the tower and nacelle in simulations to accurately model tip vortex breakdown and its significance for wake recovery. Note that conditional sampling for misalignment angle is not applied for this analysis because of a limited amount of 30-second periods that fit the aligned criteria. Therefore, we expect that the trends of the frequency peaks may become clearer when conditional sampling based on yaw error is implemented with more data.

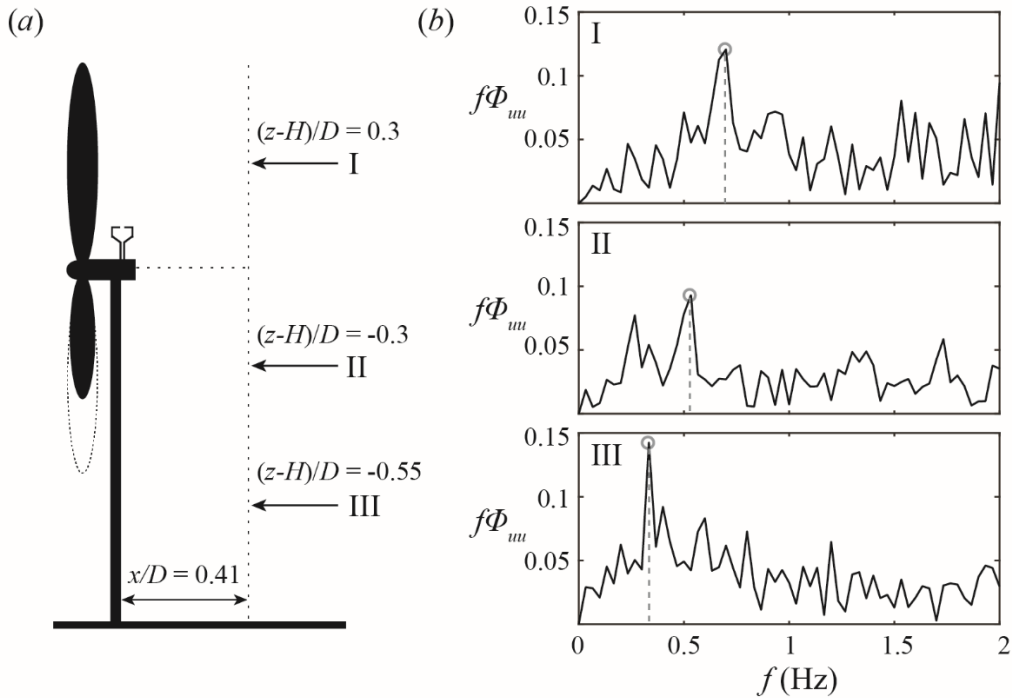

FIGURE 11 (*a*) Schematic illustrating the locations in the wake where the streamwise velocity spectra are calculated. (*b*) Sample premultiplied spectra at each wake location – I: Above the hub and below the top blade tip, II: Above the bottom blade tip and below the hub, and III: Below the bottom blade tip. The location of the peak in each spectrum is indicated by a circle and a vertical dashed line.

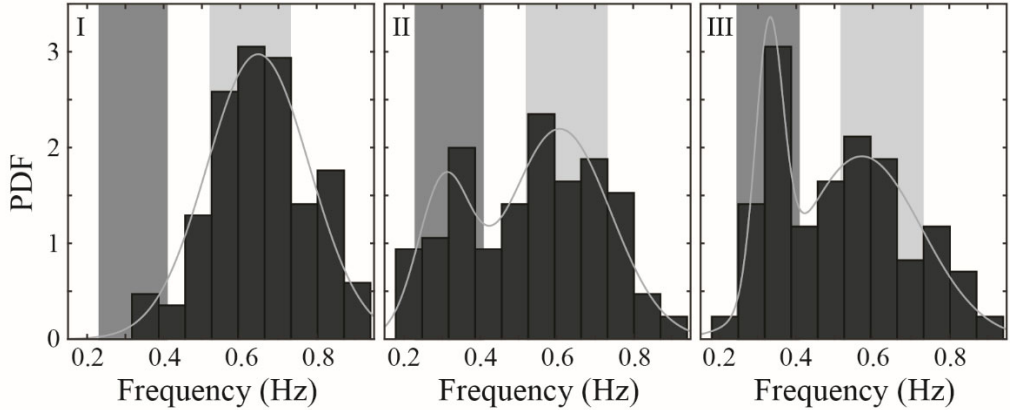

FIGURE 12 PDFs of the frequencies at each of the three wake locations described in figure 11 fit with bimodal normal distributions (light gray line). The dark gray and light gray bars indicate the frequency regions corresponding to tower vortex shedding and blade pass frequency, respectively.

## 4. Conclusions and discussion

### 4.1. *Conclusions*

This study presents the first investigation of the effect of the nacelle and tower on the near wake of a utility-scale wind turbine using super-large-scale particle image velocimetry (SLPIV) and flow visualization with natural snow. The mean flow statistics of the turbulent wake flow field is first examined using 62 minutes of SLPIV data from March 12$^{th}$, 2017, with a field of view (FOV) of 125 m (vertical) × 70 m (streamwise) centred $0.41D$ downstream of the turbine tower and approximately in the central tower plane ($0.06D$ spanwise offset). The time-averaged wake velocity field exhibits a region of accelerated flow around the hub region caused by the reduction in axial induction at the blade roots. Correspondingly, the velocity deficit profile forms a clear double-Gaussian shape. The in-plane turbulent kinetic energy (TKE) shows maxima at regions of strong shear, i.e., the blade tips and hub, and a reduction behind the tower caused by the tower breaking up large-scale streamwise structures that are responsible for turbulence production in the boundary layer.

Snow voids reveal coherent structures shed from the blades, hub, and tower. A distinct difference in flow pattern is observed above and below the hub due to the strong interaction between the tower and blade-generated vortices below the hub. The hub vortex, distinct from the blade root vortices, is visible in the region directly behind the hub. The hub wake meandering frequency is quantified and shown to match the frequency of vortex shedding from an Ahmed body ($St = 0.06$). Persistent hub wake deflection is also observed and found to be strongly connected with the turbine yaw error. Finally, the temporal characteristics of the tower-affected wake below the nacelle are quantified. The co-presence of two dominant frequencies is observed, one corresponding to blade vortex shedding at the blade pass frequency and the other corresponding to tower vortex shedding at $St = 0.2$. The dominance of each frequency varies with elevation.

### 4.2. *Discussion*

The high-resolution field-scale data presented in this study highlights the significance of the development of the hub and tower structures for the near-wake flow field. Such information should be incorporated into near-wake models to improve our prediction of wake growth. Specifically, the acceleration of the flow around the turbine nacelle provides support for the double-Gaussian near-wake model proposed by Keane *et al.* (2016). The significant change of TKE distribution due to the presence of turbine nacelle and tower reinforces the importance of their inclusion in numerical models of wake development. Specifically, to accurately represent the wake turbulence, the effect of nacelle and tower may be considered as extra forcing terms in actuator-type models, which could lead to enhanced mixing and faster wake recovery. Additionally, our study suggests that the factors influencing hub wake meandering may be different for field scale turbines than model scale turbines where the nacelle size is larger relative to the rotor diameter. Our field measurement suggests that the hub wake meandering in the near wake may be more associated with the vortex shedding from the nacelle structure rather than the interaction with vortices shed from the blades, which could become more significant as the hub wake expands downstream. The hub wake is also significantly influenced by yaw error, suggesting the need to consider the yawed condition of the turbine in the modelling of hub wake behavior. The factors influencing hub wake behavior are especially important because the hub wake has been shown to significantly influence downstream whole wake meandering in laboratory-scale studies (Kang *et al*. 2014, Howard *et al*. 2015, Foti *et al*. 2016), whereas its influence at the field scale is not yet fully understood. Our study has also demonstrated the strong influence of the tower wake on the near-wake velocity field and the TKE in the bottom half of the turbine wake (i.e. below the hub). According to an LES study from Santoni *et al*. (2017) comparing the wake of a rotor with and without a tower, such influence can lead to modified far wake development and wake recovery in comparison to non-tower cases. Particularly, the influence on the bottom half of the turbine wake can potentially impact the interaction of the turbine wake with the ground. This may further affect how the turbine wake changes the heat flux and surface temperature by bringing warmer or cooler air from the upper part of the boundary layer to the surface as noted in Roy & Traiteur (2010), or changes the fluxes of $CO_2$ (in addition to heat flux) that impacts the growth of crops (Rajewski *et al*. 2013). With significantly higher spatial and temporal resolution, our measurements are able to provide detailed flow features that are not available from other state-of-the-art field measurement techniques (e.g., lidar, sodar, radar, etc.). Our study can provide a link between these features and the large-scale wake behavior investigated in other field scale studies, elucidating the mechanisms that influence wake development under given conditions. By linking the measurement of near-wake structures with far-wake measurement, wake growth and recovery can be predicted more accurately. Additionally, our study can inform laboratory-scale experiment design. The results of our study have reinforced the importance of accurately modelling the field-scale turbine geometry including accurately scaling the nacelle and tower, and using blade profiles that capture the axial induction reduction near the hub.

We acknowledge our results are limited by the inherent variability of field data and the restrictions it places on analysis techniques (e.g., conditional sampling, spectral analysis, etc.). Moreover, it is worth noting that the present data is limited to a single set of atmospheric conditions, while changes in these conditions such as wind speed, boundary layer stability, and atmospheric turbulence have been shown to significantly influence wake behavior by laboratory scale and LES studies (Chamorro & Porté-Agel 2009, Chamorro & Porté-Agel 2010, Wu & Porté-Agel 2012). Particularly, some of variabilities of the hub and tower flow structures observed in our current measurements are connected with the effect of coherent structures in the incoming atmospheric boundary layer. Such influence of incoming conditions could be assessed in future studies by simultaneously capturing incoming and near-wake flow fields with SLPIV. Finally, previous studies have shown that turbine operation (e.g. changes in blade pitch, Dasari *et al.* 2019) also affects whole wake and blade-generated coherent structure behavior, so future studies can continue to explore the influence of turbine operation on tower and hub-induced structures.

## Acknowledgements

This work was supported by the National Science Foundation CAREER award (NSF-CBET-1454259), Xcel Energy through the Renewable Development Fund (grant RD4-13) as well as IonE of University of Minnesota. We also thank the faculties and engineers from St Anthony Falls Laboratory, including M. Guala, S. Riley, J. Tucker, C. Ellis, J. Marr, C. Milliren and D. Christopher for their assistance in the experiments.

## REFERENCES

BARTHELMIE, R. J., FRANDSEN, S. T., NIELSEN, M. N., PRYOR, S. C., RETHORE, P. E. & JØRGENSEN, H. E. 2007 Modelling and measurements of power losses and turbulence intensity in wind turbine wakes at middelgrunden offshore wind farm. *Wind Energy* **10** (6), 517–528.

BARTHELMIE, R. J., HANSEN, K., FRANDSEN, S. T., RATHMANN, O., SCHEPERS, J. G., SCHLEZ, W., PHILLIPS, J., RADOS, K., ZERVOS, A., POLITIS, E. S. & CHAVIAROPOULOS, P. K. 2009 Modelling and measuring flow and wind turbine wakes in large wind farms offshore. *Wind Energy* **12** (5), 431–444.

CASTRO, I. P. & ROBINS, A. G. 1977 The flow around a surface-mounted cube in uniform and turbulent streams. *J. Fluid Mech.* **79**, 307-335.
CHAMORRO, L. P. & PORTÉ-AGEL, F. 2009 A wind-tunnel investigation of wind-turbine wakes: boundary-layer turbulence effects. *Boundary-Layer Meteorol.* **132** (1), 129-149.
CHAMORRO, L. P. & PORTÉ-AGEL, F. 2010 Effects of thermal stability and incoming boundary-layer flow characteristics on wind-turbine wakes: a wind-tunnel study. *Boundary-Layer Meteorol.* **136** (3), 515-533.
DASARI, T., WU, Y., LIU, Y. & HONG, J. 2019 Near-wake behaviour of a utility-scale wind turbine. *J. Fluid Mech.* **859**, 204-246.
DUELL, E. G. & GEORGE, A. R. 1999 Experimental study of a ground vehicle body unsteady near wake. *SAE Transactions* **108** (6), 1589-1602.
FELLI, M., CAMUSSI, R. & DI FELICE, F. 2011 Mechanisms of evolution of the propeller wake in the transition and far fields. *J. Fluid Mech.* **682**, 5-53.
FOTI, D., YANG, X., CAMPAGNOLO, F., MANIACI, D. & SOTIROPOULOS, F. 2018 Wake meandering of a model wind turbine operating in two different regimes. *Phys. Rev. Fluids* **3** (5), 1-34.
FOTI, D., YANG, X., GUALA, M. & SOTIROPOULOS, F. 2016 Wake meandering statistics of a model wind turbine: Insights gained by large eddy simulations. *Phys. Rev. Fluids* **1** (4), 044407.
GALLACHER, D. & MORE, G. 2014 Lidar measurements and visualisation of turbulence and wake decay length. *Alpha Ventus AV07 Lidar Programme.* SgurrEnergy Ltd.
GÖÇMEN, T., LAAN, P. VAN DER, RÉTHORÉ, P. E., DIAZ, A. P., LARSEN, G. C. & OTT, S. 2016 Wind turbine wake models developed at the technical university of Denmark: A review. *Renew. Sustain. Energy Rev.* **60**, 752–769.
GOHLKE, M., BEAUDOIN, J. F., AMIELH, M. & ANSELMET, F. 2007 Experimental analysis of flow structures and forces on a 3D-bluff-body in constant cross-wind. *Exp. Fluids* **43** (4), 579-594.
HAMILTON, J. M., KIM, J. & WALEFFE, F. 1995 Regeneration mechanisms of near-wall turbulence structures. *J. Fluid Mech.* **287**, 317-348.
HANCOCK, P. E. & PASCHEKE, F. 2014 Wind-tunnel simulation of the wake of a large wind turbine in a stable boundary layer: Part 2. the wake flow. *Boundary-Layer Meteorol.* **151** (1), 23-37.
HAND, M. M., SIMMS, D. A., FINGERSH, L. J., JAGER, D. W., COTRELL, J. R., SCHRECK, S. & LARWOOD, S. M. 2011 Unsteady aerodynamics experiment phase VI: wind tunnel test configurations and available data campaigns (No. NREL/TP-500-29955). National Renewable Energy Laboratory.
HONG, J., TOLOUI, M., CHAMORRO, L. P., GUALA, M., HOWARD, K., RILEY, S., TUCKER, J. & SOTIROPOULOS, F. 2014 Natural snowfall reveals large-scale flow structures in the wake of a 2.5-MW wind turbine. *Nat. Commun.* **5**, 4216.
HOWARD, K. B., SINGH, A., SOTIROPOULOS, F., & GUALA, M. 2015 On the statistics of wind turbine wake meandering: An experimental investigation. *Phys. Fluids* **27** (7), 075103.
HSU, M.-C., AKKERMAN, I. & BAZILEVS, Y. 2014 Finite element simulation of wind turbine aerodynamics: validation study using NREL Phase VI experiment. *Wind Energy* **17**, 461-481.
IUNGO, G. V., VIOLA, F., CAMARRI, S., PORTÉ-AGEL, F. & GALLAIRE, F. 2013 Linear stability analysis of wind turbine wakes performed on wind tunnel measurements. *J. Fluid Mech.* **737**, 499-526.
IUNGO, G. V., WU, Y.T. & PORTÉ-AGEL, F. 2012 Field measurements of wind turbine wakes with lidars. *J. Atmos. Ocean. Technol.* **30** (2), 274-287.
IVANELL, S., MIKKELSEN, R., SØRENSEN, J. N. & HENNINGSON, D. 2010 Stability analysis of the tip vortices of a wind turbine. *Wind Energy* **13** (8), 705–715.
JACOBI, I. & MCKEON, B. J. 2011 New perspectives on the impulsive roughness-perturbation of a turbulent boundary layer. *J. Fluid Mech.* **677**, 179-203.
JIMÉNEZ, J. & PINELLI, A. 1999 The autonomous cycle of near-wall turbulence. *J. Fluid Mech.* **389**, 335-359.
KANG, S., YANG, X. L. & SOTIROPOULOS, F. 2014 On the onset of wake meandering for an axial flow turbine in a turbulent open channel flow. *J. Fluid Mech.* **744**, 376–403.
KÄSLER, Y., RAHM, S., SIMMET, R. & KÜHN, M. 2010 Wake measurements of a multi-MW wind turbine with coherent long-range pulsed Doppler wind lidar. *J. Atmos. Ocean. Technol.* **27** (9), 1529-1532.
KEANE, A., AGUIRRE, P. E. O., FERCHLAND, H., CLIVE, P. & GALLACHER, D. 2017 An analytical model for a full wind turbine wake. *J. Phys.: Conf. Ser.* **753** (3), 032039.
KEOGH, J., BARBER, T., DIASINOS, S. & DOIG, G. 2016 The aerodynamic effects on a cornering Ahmed body.

*J. Wind Engng Ind. Aerodyn.* **154**, 34-46.

KRAJNOVIĆ, S. 2011 Flow around a tall finite cylinder explored by large eddy simulation. *J. Fluid Mech.* **676**, 294-317.

KRAJNOVIĆ, S. & DAVIDSON, L. 2003. Numerical study of the flow around a bus-shaped body. *J. Fluids Eng.*, **125** (3), 500-509.

LI, Y., PAIK, K. J., XING, T. & CARRICA, P. M. 2012 Dynamic overset CFD simulations of wind turbine aerodynamics. *Renew. Energy* **37** (1), 285-298.

LIGNAROLO, L. E. M., RAGNI, D., SCARANO, F., SIMÃO FERREIRA, C. J., & VAN BUSSEL, G. J. W. 2015 Tip-vortex instability and turbulent mixing in wind-turbine wakes. *J. Fluid Mech.* **781**, 467-493.

LYNCH, C. E. & SMITH, M. J. 2013 Unstructured overset incompressible computational fluid dynamics for unsteady wind turbine simulations. *Wind Energy* **16** (7), 1033-1048.

MAGNUSSON, M. 1999 Near-wake behaviour of wind turbines. *J. Wind Engng Ind. Aerodyn.* **80** (1-2), 147-167.

NEMES, A., JACONO, D. LO, BLACKBURN, H. M. & SHERIDAN, J. 2015 Mutual inductance of two helical vortices. *J. Fluid Mech.* **774**, 298–310.

OKULOV, V. L. & SØRENSEN, J. N. 2007 Stability of helical tip vortices in a rotor far wake. *J. Fluid Mech.* **576**, 1–25.

PARK, C. W. & LEE, S. J. 2002 Flow structure around a finite circular cylinder embedded in various atmospheric boundary layers. *Fluid Dyn. Res.* **30** (4), 197–215.

PATHOKONDA, G. 2013 Structure of turbulent channel flow perturbed by cylindrical roughness elements. M.Sc. thesis, Department of Aerospace Engineering, University of Illinois at Urbana-Champaign.

PATHIKONDA, G. & CHRISTENSEN, K. T. 2015 Structure of turbulent channel flow perturbed by a wall-mounted cylindrical element. *AIAA J.* **53** (5), 1277-1286.

PEARSON, B., ELAVARASAN, R. & ANTONIA, R. 1998 Effect of a short roughness strip on a turbulent boundary layer. *Appl. Sci. Res.* **59**, 61-75.

RAJEWSKI, D. A., TAKLE, E. S., LUNDQUIST, J. K., ONCLEY, S., PRUEGER, J. H., HORST, T. W., RHODES, M. E., PFEIFFER, R., HATFIELD, J. L., SPOTH, K. K. & DOORENBOS, R. K. 2013 Crop wind energy experiment (CWEX): observations of surface-layer, boundary layer, and mesoscale interactions with a wind farm. *Bull. Amer. Meteor. Soc.* **94** (5), 655-672.

ROY, S. B. & TRAITEUR, J. J. 2010 Impacts of wind farms on surface air temperatures. *PNAS* **107** (42), 17899-17904.

RYAN, M. D., ORTIZ-DUEÑAS, C. & LONGMIRE, E. K. 2011 Effects of simple wall-mounted cylinder arrangements on a turbulent boundary layer. *AIAA J.* **49** (10), 2210-2220.

SANTONI, C., CARRASQUILLO, K., ARENAS-NAVARRO, I. & LEONARDI, S. 2017 Effect of tower and nacelle on the flow past a wind turbine. *Wind Energy* **20** (12), 1927-1939.

SARMAST, S., DADFAR, R., MIKKELSEN, R. F., SCHLATTER, P., IVANELL, S., SØRENSEN, J. N. & HENNINGSON, D. S. 2014 Mutual inductance instability of the tip vortices behind a wind turbine. *J. Fluid Mech.* **755**, 705–731.

SCHULZ, C., LETZGUS, P. LUTZ, T. & KRÄMER, E. 2017 CFD study on the impact of yawed inflow on loads, power and near wake of a generic wind turbine. *Wind Energy* **20** (2), 253-268.

SHERRY, M., SHERIDAN, J. & LO JACONO, D. 2013 Characterisation of a horizontal axis wind turbine's tip and root vortices. *Exp. Fluids* **54** (3), 1417.

SHIH, W. C. L., WANG, C., COLES, D. & ROSHKO, A. 1993 Experiments on flow past rough circular cylinders at large Reynolds numbers. *J. Wind Engng. Ind. Aerodyn.* **49** (1-3), 351-368.

SIMMS, D., SCHRECK, S., HAND, M. & FINGERSH, L.J. 2001 NREL unsteady aerodynamics experiment in the NASA-Ames wind tunnel: a comparison of predictions to measurements (No. NREL/TP-500-29494). National Renewable Energy Laboratory.

SØRENSEN, J. N. 2011 Instability of helical tip vortices in rotor wakes. *J. Fluid Mech.* **682**, 1–4.

STOESSER, G. P. T., FRÖHLICH, J., KAPPLER, M. & RODI, W. 2010 Large eddy simulations and experiments of flow around finite-height cylinders. *Flow, Turbul. Combust.* **84** (2), 239-275.

SUMNER, D., HESELTINE, J. L., & DANSEREAU, O. J. P. 2004 Wake structure of a finite circular cylinder of small aspect ratio. *Exp. Fluids* **37** (5), 720-730.

TOLOUI, M., RILEY, S., HONG, J., HOWARD, K., CHAMORRO, L. P., GUALA, M. & TUCKER, J. 2014 Measurement of atmospheric boundary layer based on super-large-scale particle image velocimetry using natural snowfall. *Exp. Fluids* **55** (5), 1737.
VERMEER, L. J., SORENSEN, J. N. & CRESPO, A. 2003 Wind turbine wake aerodynamics. *Prog. Aerosp. Sci.* **39** (6–7), 467–510.
VIOLA, F., IUNGO, G. V., CAMARRI, S., PORTÉ-AGEL, F. & GALLAIRE, F. 2014 Prediction of the hub vortex instability in a wind turbine wake: stability analysis with eddy-viscosity models calibrated on wind tunnel data. *J. Fluid Mech.* **750**, R1.
WANG, J., MCLEAN, D., CAMPAGNOLO, F., YU, T. & BOTTASSO, C. L. 2017 Large-eddy simulation of waked turbines in a scaled wind farm facility. *J. Phys.: Conf. Ser.* **854**, 012047.
WANG, Q., ZHOU, H. & WAN, D. 2012 Numerical simulation of wind turbine blade-tower interaction. *J. Mar. Sci. Appl.* **11** (3), 321-327.
WHALE, J., PAPADOPOULOS, K. H., ANDERSON, C. G., HELMIS, C. G. & SKYNER, D. J. 1996 A study of the near wake structure of a wind turbine comparing measurements from laboratory and full-scale experiments. *Sol. Energy* **56** (6), 621-633.
WIDNALL, S. E. 1972 The stability of a helical vortex filament. *J. Fluid Mech.* **54** (4), 641–663.
WU, Y. T. & PORTÉ-AGEL, F. 2012 Atmospheric turbulence effects on wind-turbine wakes: An LES study. *Energies* **5** (12), 5340-5362.
YANG, X., HOWARD, K. B., GUALA, M. & SOTIROPOULOS, F. 2015 Effects of a three-dimensional hill on the wake characteristics of a model wind turbine. *Phys. Fluids* **27** (2), 025103.
YANG, X. & SOTIROPOULOS, F. 2018 A new class of actuator surface models for wind turbines. *Wind Energy* **21** (5), 285-302.
ZAHLE, F., SØRENSEN, N. N., & JOHANSEN, J. 2009 Wind turbine rotor-tower interaction using an incompressible overset grid method. *Wind Energy* **12** (6), 594-619.
ZHANG, W., MARKFORT, C.D. & PORTÉ-AGEL, F. 2012 Near-wake flow structure downwind of a wind turbine in a turbulent boundary layer. *Exp. Fluids* **52** (5), 1219-1235.